\documentclass[preprint2]{emulateapj}

\usepackage{natbib}  
\usepackage{amssymb}

\usepackage{comment}
\usepackage{graphicx}

\bibpunct{(}{)}{;}{a}{}{,} 

\def\cm{\,{\rm cm}}

\def\ergscm2 {erg\,s$^{-1}$cm$^{-2}$}

\def\cm2 {cm$^{-2}$}

\def\aap {A\&A}
\def\apj {ApJ}

\def\wdsrc {G87$-$7}
\def\grbsrc {GRB 110328A}

\shorttitle{\Large Quark-Novae in LMXBs II:
Application to  \wdsrc~  and to \grbsrc}
\shortauthors{Ouyed et al.}

\begin{document}

\title{ Quark-Novae in  Low-mass X-ray Binaries II:\\
 Application to \wdsrc~  and to \grbsrc}

\author{Rachid Ouyed}
\affil{Department of Physics and Astronomy, University of Calgary, 
2500 University Drive NW, Calgary, Alberta, T2N 1N4 Canada}
\author{Jan Staff}
\affil{Department of Physics and Astronomy, Louisiana State University,
202 Nicholson Hall, Tower Dr., Baton Rouge, LA 70803-4001, USA}
\and
\author{Prashanth Jaikumar}
\affil{Department of Physics and Astronomy, California State University  Long Beach,  1250 Bellflower Blvd., Long Beach CA 90840}

\begin{abstract}

We propose a simple model explaining two outstanding astrophysical problems related to compact objects: (1) that of stars such as G87$-$7 (alias EG 50) that constitute a class of relatively low-mass white dwarfs which nevertheless fall away from the C/O composition and (2) that of GRB 110328A/Swift J164449.3$+$57345  which showed spectacularly long-lived strong X-ray flaring, posing a challenge to standard  GRB models. We argue that both these observations may have an explanation within the unified framework of a Quark-Nova occurring in a low-mass X-ray binary (neutron star-white dwarf). For LMXBs where the binary separation is sufficiently tight, ejecta from the exploding Neutron Star triggers nuclear burning in the white dwarf on impact, possibly leading to Fe-rich composition compact white dwarfs with mass $0.43M_{\odot}<M_{\rm WD} < 0.72M_{\odot}$, reminiscent of \wdsrc.  Our results rely on the assumption, which ultimately needs to be tested by hydrodynamic and nucleosynthesis simulations, that under
certain circumstances the WD can avoid the thermonuclear runaway.  For  heavier white dwarfs (i.e. $M_{\rm WD} > 0.72M_{\odot}$) experiencing the QN shock, degeneracy will not be lifted when Carbon burning begins, and a sub-Chandrasekhar Type Ia Supernovae may result in our model.  Under slightly different conditions, and for pure He white dwarfs (i.e. $M_{\rm WD} < 0.43M_{\odot}$), the white dwarf is ablated and its ashes raining down on the Quark star leads to accretion-driven X-ray luminosity with energetics and duration reminiscent of \grbsrc.  We predict additional flaring activity towards the end of the accretion phase if the Quark star turns into a Black Hole.

\end{abstract}

\section{Introduction}

In a recent paper by \cite{osj}, hereafter OSJ, we presented a detailed
account of the Quark-Nova model as applied to short GRBs. The jist of this model is that a Quark-Nova (QN), which is an explosive transition of a Neutron Star to a Quark Star (QS), when it occurs in a low-mass X-ray binary (LMXBs; see \S \ref{sec:configuration} below), can heat and ablate the white dwarf (WD) companion leading to the observed extended emission (EE) in some short GRBs, while prompt emission arises from shocking a circumbinary disk. 
In this work, we use aspects of the same (OSJ) model to explain the long-lived X-ray flaring seen in GRB 110328A and in the process find a possible resolution  for a longstanding puzzle regarding WD compositions. The puzzle is this: There is at present no clear scenario that can explain the formation of a low mass WD with a heavy core composition, including the possibility of an iron core. Provencal et al. (1998) found a handful of such stars using Hipparcos parallax data which do not fall on the expected C/O relationships for average mass WDs. This includes the most outstanding case, that of G 87$-$7 (alias EG 50) which appears to fall instead right on the Hamada \& Salpeter (1961) zero-temperature curve for Fe in the M-R plane (see Figure 3 of Provencal et al. 1998). 

Before we present numerical estimates to support our arguments below, we emphasize that the crux of our reasoning can be stated simply: For tight binary separation at the moment the QN happens, ejecta from the Neutron star (NS) impacts the WD leading to burning of the C/O WD up to Iron, while lifting
the degeneracy, leading to stable but ``anomalous" WD compositions instead of
a Type I supernova. If the WD is light enough, thermal ablation and subsequent accretion of WD material onto the Quark Star occurs. This leads to an accretion-powered X-ray luminosity  distinguished by features (energetics, timescale and temporal variability) reminiscent of \grbsrc \,(\cite{cummings11}).

The paper is organized as follows: In \S 2, we describe the binary configuration
 in our model and estimate the conditions that determine the fate of the WD (ablated or not). In \S 3, we apply our
model to G87-7 and in \S 4 to GRB 110328A. In both cases, we discuss predictions of our model that can be tested. We conclude in \S 5.

  \section{Quark-Nova in a LMXB}
  
  When matter in the core of a NS reaches the quark deconfinement density, a phase transition to quark matter can occur. Preliminary numerical
  simulations hint at potential instabilities that can trigger a detonative
  transition (Niebergal et al. 2011). Ouyed et al. (2002) termed this a ``Quark-Nova" and its likely outcome is the ejection of the outermost layers of the NS, with kinetic energy of the relativistic QN ejecta $E_{\rm QN}^{\rm K}=  (\Gamma_{\rm QN}-1) M_{\rm QN}c^2 \sim 10^{52}$ ergs for typical  ejected mass of $M_{\rm QN}\sim 10^{-3}M_{\odot}$ and  ejecta's Lorentz factor $\Gamma_{\rm QN}\sim 10$  (see Ker\"anen et al. (2005)). 
  
  \subsection{The Binary Configuration}
  \label{sec:configuration}
  
  In our model,  a QN would form after the first accretion phase,
when the increase in central density is driven by
spin-down and/or gravitational wave emission
(eg., \citep{staff06}; see also \cite{staff11}
for spin-down induced by gravitational waves). 
In fact, the QN has a higher probability of occurrence after the first accretion
phase since the mass of the NS in such a case would be higher.
Furthermore, gravitational wave emission can spin the star down faster
than magnetic braking \citep{staff11},
causing a large change in the central density.
We also note that, in principle, the QN could also occur
in the second accretion phase after the WD has reached its Roche Lobe (RL).
In summary, the QN could occur under three situations in these systems,
depending on the initial mass of the neutron star:
 (i) During the first accretion phase; (ii) From  spin-down,  caused by a combination of
  gravitational waves and magnetic braking,  following the first accretion phase;
 (iii) In the second accretion phase.  
  
In the scenario (iii), the (C-O or He) WD would fill its RL
and the second accretion phase starts. This scenario is reminiscent
 of Ultracompact X-ray binaries (UCXBs). These 
 are binaries with orbital periods
shorter than  $\sim$ 1 hr and are believed to be 
a subset of the LMXBs.
Evidence was found for carbon/oxygen as well as helium/nitrogen discs 
 and no evidence for (traces of) hydrogen in some UCXBs
 (e.g. \citep{nelemans06}  for a discussion);
 This suggests that the donors in the observed systems are degenerate WDs.
 In some of these, only Carbon and Oxygen lines were identified. 
 which supports the notion of C-O WD donors. (e.g \citep{juet01}). 
Recent studies concludes that 
  an important  fraction of LMXBs may be ultracompact
  (e.g. \citep{zand05}).
Obviously a thorough population synthesis is needed to address 
the contribution to UCXBs from systems with non-degenerate,
semi-degenerate, and degenerate donors. In general, we favour scenarios/options
(ii) and/or (iii)  in our model.
  
If a QN happens in a binary system with a WD and a NS orbiting each
other, what will be the fate of the WD?

\subsection{Fate of the White Dwarf}

In the paper OSJ, we found that depending on the binary separation, $a$,
the impact of the ejecta with the WD triggers various nuclear burning processes in the latter. The temperature per nucleon of the shocked WD is
\begin{equation}
\label{eq:Textend}
k_{\rm B} T_{\rm WD} \sim 17.6 \ {\rm keV} \frac{\mu_{\rm WD, 2} E_{\rm QN, 52}^{\rm KE}}{a_{10}^2 M_{\rm WD,
0.43}^{5/3}}\ ,
\end{equation}
where $\mu_{\rm WD}$ is the mean molecular weight in units of 2
(for a CO WD), the WD mass is in units of $0.43M_{\odot}$ (chosen
a posteriori since 0.43$M_{\odot}$ turns out to be the border between ablation and survival of the WD) and the binary separation $a_{10}$ is in units of $10^{10}$ cm. Considering CO WDs, we see that for carbon-carbon burning to start (\cite{Lang}), the temperature must exceed $\sim 70\ {\rm keV}$. This implies:

\begin{equation}
a_{10}^2M_{\rm WD,0.43}^{5/3} < 0.25 \mu_{\rm WD, 2} E_{\rm QN,52}^{\rm KE},
\label{eq:condition1}
\end{equation}
or in other words the separation must be $a_{10} < 0.5$ in order for the
temperature to be sufficiently high that carbon-carbon burning can occur in
a $0.43 M_\odot$ CO WD. We note that this is still larger than the
separation at which the WD begins to overflow its Roche lobe:
\begin{equation}
a_{\rm RL}\sim9.2\times10^9\bigg(\frac{M_{\rm
T}}{M_\odot}\bigg)^{2/3}\bigg(\frac{0.1M_\odot}{M_{\rm WD}}\bigg)
\label{eq:rochelobe}
\end{equation}
which is $3.7\times10^9$ cm for a $0.43 M_\odot$ WD and a $1.8 M_\odot$ NS ($M_T=M_{\rm NS}+M_{\rm WD}$).
     
Nuclear burning results in a total energy release\footnote{The number of baryons in a $0.43M_{\odot}$ WD is
$n_{\rm baryons}= M_{\rm WD}/m_{\rm H}\simeq 5.1\times10^{56} M_{\rm
WD,0.43}$. Nuclear burning releases 
$E_{\rm nuc.}=n_{\rm baryons}\times \Delta \simeq 
4.1\times10^{50} {\rm erg}~M_{\rm WD,0.43} \Delta_{0.5}$ where
the energy released per baryon is parametrized by $\Delta_{0.5}$ given
in units of 0.5 MeV. If $10\%$ ($\eta=0.1$) of this energy is converted to kinetic energy
 (i.e. $0.5 M_{\rm WD}v_{\rm nuc., ejec}^2 = \eta E_{\rm nuc.}$) then, the WD nuclei gain a speed of:
$v_{\rm nuc., ejec} \simeq 3.1\times10^3~{\rm km~s^{-1}} 
\eta_{0.1}^{1/2}\Delta_{0.5}^{1/2}$.} $\sim  4\times 10^{50}
\times M_{\rm  WD,  0.43}
\Delta_{0.5}$ ergs, where $\Delta_{0.5}$ is the
energy gained per baryon due to nuclear burning in terms of $0.5\,{\rm
MeV/baryon}$. We estimate that the processed nuclei are then ejected at speeds
 \begin{equation}
 v_{\rm nuc., ejec.}\simeq  3.1\times 10^3\
{\rm km\ s}^{-1}\ {\eta_{0.1}}^{1/2} \Delta_{0.5}^{1/2}\,\,,
\end{equation}
where $\eta$ is the nuclear-to-kinetic-energy conversion efficiency,
taken to have a fiducial value of 0.1. The WD escape speed is 
$v_{\rm WD, esc.}\simeq 3.2 \times 10^3\ {\rm km s}^{-1}  
\times M_{\rm WD, 0.43}^{2/3}$ using the WD equation of state as given in 
eq.(2) of OSJ. Thus, the condition for the processed material to escape the 
WD (ablation) $v_{\rm nuc., ejec.} > v_{\rm WD, esc.}$ implies
\begin{equation}
\label{eq:condition2}
M_{\rm WD} < 0.43 M_{\odot} \ {\eta_{0.1}}^{3/4} \Delta_{0.5}^{3/4} \,\,.
\end{equation}

The sufficient condition for the WD to burn {\it and} be ablated  is 
given by combining eq.(\ref{eq:condition1}) and eq.(\ref{eq:condition2}).
Note that Eq.(\ref{eq:condition2}) provides a necessary condition for ablation of the WD mass (for a given $\eta$
and $\Delta$). 
We can write the condition in eq.(\ref{eq:condition1}) in terms of a maximum binary separation for a given WD mass:  
\begin{equation}
\label{amax}
a_{10} < a_{10}^{\rm max}=0.5 \frac{\mu_{\rm WD, 2}^{1/2}{E_{\rm QN,
52}^{\rm KE}}^{1/2}}{M_{\rm WD,0.43}^{5/6}}\ . 
\end{equation}
The condition for the WD to burn {\it without} ablation follows simply by 
reversing the inequality in eq.(\ref{eq:condition2}) while retaining 
eq.(\ref{eq:condition1})
\begin{equation}
\label{eq:condition-check}
0.43 M_{\odot}~\eta_{0.1}^{3/4}~\Delta_{0.5}^{3/4} < M_{\rm WD} <  0.18 M_{\odot}
\frac{\mu_{\rm WD, 2}^{3/5}{E_{\rm QN, 52}^{\rm KE}}^{3/5}}{a_{10}^{6/5}}\ .
\end{equation}
From the consistency check that the upper bound in eq.(\ref{eq:condition-check}) exceeds the lower bound, 
we obtain $a_{10} < 0.5$ for fiducial 
values\footnote{This bound on $a_{10}$ is the same as $a_{10}^{\rm max}$ since 
checking the consistency of the bounds in eq.(\ref{eq:condition-check}) is 
the same as saturating the bound in eq.(\ref{eq:condition2}).} of 
$\mu_{\rm WD}, E_{\rm QN}^{KE}, \eta, \Delta$ for a CO WD (implying that
$\mu_{\rm WD}=2$).  The resulting WD should be 
rich in  elements heavier than C and O (see \S \ref{sec:g87-7-sub1}).
 The fate of the WD ashes in the ablation case is explored in \S \ref{sec:grb11}. 
If the QN happens in a LMXBs with $M_{\rm WD} > 0.43M_{\odot}$ when the binary 
separation is $a > a_{10}^{\rm max}$, it will
 not lead to nuclear burning in the WD, but such a system should have 
distinct signatures as the heated WD (with $kT_{\rm WD} < 70$ keV) should 
cool in softer X-rays compared to the case of tight binaries. The different 
regimes are summarized in Table 1 and applied to particular observational 
cases in the following sections. 

\subsection{Pure He WD}

A CO WD can be heavy enough to escape ablation, but we show below that a pure He WD most likely cannot. For a pure He WD, $\mu_{\rm WD} = 4/3$, and its temperature must exceed $8.6$ keV in order to start burning Helium in the triple-alpha process. This implies that

\begin{equation}
a_{10}^2 M_{\rm WD,0.25}^{5/3} < 3.37\mu_{\rm WD,4/3}E_{\rm QN, 52}^{\rm KE} \ ,
\end{equation}
where the WD mass is units of $0.25M_{\odot}$.

Nuclear burning of Helium through the triple-$\alpha$ process 
results in a total energy release of $\sim 2.4\times10^{50}\times M_{\rm WD,0.25}\Delta_{0.5}$ ergs. This leads to the same
nuclei speed as before, since that is independent of the WD mass:

\begin{equation}
 v_{\rm nuc., ejec.}\simeq  3.1\times 10^3\
{\rm km\ s}^{-1}\ {\eta_{0.1}}^{1/2} \Delta_{0.5}^{1/2}\,\,;
\end{equation}

hence Ineq.~\ref{eq:condition2} also remains the same.
However, the mass of a pure He WD is unlikely to be as large as $0.43 M_\odot \eta_{0.1}^{3/4} \Delta_{0.5}^{3/4}$ unless $\eta$ or $\Delta$ are much less than their fiducial values of 0.1 and 0.5 respectively. Hence we expect a pure He WD to ablate in most cases. The maximum binary separation that allows for 
burning of the He WD is:

\begin{equation}
a_{10}<a_{10}^{\rm max}=1.84 \frac{\mu_{\rm WD,4/3}E_{\rm QN,52}^{\rm
KE}}{M_{\rm WD,0.25}^{5/6}}.
\end{equation}

For completeness, although this is unlikely to happen for a pure He WD, we write down the condition for burning without ablation:

\begin{equation}
0.43M_\odot \eta_{0.1}^{3/4}~\Delta_{0.5}^{3/4} < M_{\rm WD} < 0.52 M_\odot 
\frac{\mu_{\rm WD,4/3}^{3/5}{E_{\rm QN,52}^{\rm KE}}^{3/5}}{a_{10}^{6/5}}.
\end{equation}

    \section{A case for \wdsrc}
    \label{sec:g87-7}

Following the conclusions of Provencal et al. (1998), the idea that \wdsrc~ could have a Fe core  subsequently gained additional support from the work of \citep{panei00}  who considered detailed evolutionary calculations based on iron core models. The fact that \wdsrc~ is not a massive star by WD standards makes it hard to understand how it could be made of iron or iron-rich material. \cite{mathews06} made a systematic study of correction terms to the WD equation of state, but none of these corrections, not even possible effects of magnetic fields can be reasonably varied to fit these compact WDs for a normal He, C, or Mg WD. 
 
Recent measurements by Hipparcos present observational evidence supporting the existence of some WD stars with iron-rich core composition.
Cores made of C, O, Ne, Mg, Si, S, and Ca appear to be excluded in favor of cores made of Ar, Ti, Cr, or Fe (or mixtures of those), and hydrogen envelopes that may be thick or thin (see Figure 2 in \cite{gilles07}).
Specifically, \cite{gilles07} concluded that the derived mass and radius for
\wdsrc~ vary in small ranges, from 0.5096 $M_{\odot}$ (0.0109 $R_{\odot}$) for the Ar core model with a thin H envelope to 0.5493 $M_{\odot}$ (0.0113 $R_{\odot}$) for the Fe core model with a thick H envelope. 
The existence of WDs consistent with an iron core relies critically on the accuracy of atmospheric parameters deduced from optical spectroscopy. e.g., follow-up and improved observations showed that a few of the exotic WDs found by Provencal et al. (1998) were in fact made of carbon-oxygen material (Provencal et al. 2002). Other objects they found show hints of Iron mixed with Carbon such as Procyon B, which, according to Provencal et al. (2002) is a rare DQZ white dwarf. There remain some candidates that are iron-rich (pending future observations) which are relevant to this work. It is our model prediction, as we show in this work, that if QNe occur in a NS-WD systems, under the right circumstances, Iron-rich low-mass WDs could form.

 Standard WD formation scenarios do not easily account for such objects (e.g. Isern, Canal, \& Labay 1991).  As best summarized in \cite{mathews06}:
{\it ``The conditions necessary to burn white-dwarf material to iron require such high densities and rapid reaction rates that it would seem impossible to fine tune the parameters of an accreting WD to avoid the thermonuclear runaway associated with a Type-Ia supernova and disruption of the star"}  (i.e. degenerate conditions prevent controlled burning and causes a thermonuclear runaway because a temperature increase does not lead to a pressure increase).

In what might be called the standard model for a SN Ia, a CO WD accretes matter until it compresses to the point that carbon is ignited just before the Chandrasekhar limit \citep{mazzali07}. Although the white dwarf may have a ``simmering" phase of order a thousand years following unstable Carbon ignition, where thermonuclear runaway is prevented by convection \citep{piro08}, ultimately, explosive burning is ignited and the WD is incinerated in seconds.

The Fermi energy of a WD close to the Chandrasekhar limit is $\sim$ 175 keV, much larger than the Carbon ignition temperature, thus burning is induced under degenerate conditions and leads to a thermonuclear runaway and the explosion. On the other hand, in our model, the small WD mass implies that the WD Fermi energy is only around 35 keV. The temperature induced by the QN shock can be comparable to or larger than this, which lifts the degeneracy as it passes through the star. We find that for WD masses $0.43 M_{\odot} < M_{\rm WD} < 0.7 M_{\odot}$, the QN shock is strong enough that $k_{\rm B} T_{\rm WD}$ exceeds the Fermi energy.

This scenario is unlike the standard near-Chandrasekhar mass WD responsible for Type Ia SN explosions, where heating does not relieve the degeneracy pressure.

Mathews et al. (2006) have proposed that the outliers found
by Provencal et al. (1998), could
be made, in part, of strange matter\footnote{Strange matter is a hypothesized
ground state of matter at high density composed of $u,d,s$ quarks. The additional degree of freedom provided by the strange quark and its negative
charge make for a thermodynamically favored state at high density, compared to
nuclear or neutron-rich matter (Bodmer (1971), Witten (1984)).}. 
Such objects would contain a tiny nucleus of strange matter and are known as strange dwarfs \citep{norm95}. Their masses fall in the approximate range $10^{-4}$ to 1 solar mass.  Structurally, they  show a much higher centrally condensed structure than an ordinary WD, and exhibit a smaller radius for a given mass than that expected from C/O cores (see \cite{matsuzaki07} for the stability
of these objects). By following the recipe provided in Mathews et al. (2006), \cite{gilles07} 
were  able to construct a  model of \wdsrc~ that satisfies both the
spectroscopic constraint imposed on the surface gravity and the
parallax constraint. They favored a strange WD with 
a mass of 0.5359 $M_{\odot}$ (7\% of which is due
to the strange matter nucleus), a radius of 0.0112 $R_{\odot}$. 
They concluded that it  is possible to interpret \wdsrc~ in a way that
does not invoke heavy element cores.

In any case, the interpretation of iron-core WDs is difficult to achieve from a conventional stellar evolution standpoint and does not fit the observed compact population within a single curve. The strange WD model requires quark matter to be somehow seeded or trapped in the core.
Below we offer an alternate explanation based on our LMXB numbers that can naturally account for Iron WD without appealing to the strange WD model.  
  
\subsection{\wdsrc~ in our model}
\label{sec:g87-7-sub1}

According to our model described in \S 2 and summarized in Table 1, a QN
occurring in a LMXB with $M_{\rm WD} > 0.43M_{\odot}$ when $a_{10} <
a_{10}^{\rm max}$ most likely leads to controlled nuclear burning of the WD
material without ablation.  The Fermi energy\footnote{The Fermi energy, $E_{\rm F}=(\hbar^2/2m_{\rm e})\times (3\pi^2n_{\rm e})^{2/3}$, is estimated by writing
 the  electron density as $n_{\rm e}=\frac{6M_{\rm WD}}{m_{\rm C}}\frac{1}{4\pi R_{\rm WD}^3/3}$; 
 $m_{\rm C}$ being the mass of a carbon atom. We adopt 
 $R_{\rm WD}=0.017 M_{\rm WD, 0.43}^{-1/3}\,(R_{\odot})$ as the radius of the WD (\cite{shapiro83}) which 
 gives $E_{\rm F}\sim 35$ keV for $M_{\rm WD}=0.43 M_\odot$.} of the WD is $\epsilon_{\rm
F}\sim {\rm 35\ keV} \times M_{\rm WD, 0.43}^{4/3}$. 
 This means that degeneracy is lifted for $0.43 M_{\odot} < M_{\rm WD} < 0.72M_{\odot}$ whenever burning conditions ($kT_{\rm WD} > 70\ {\rm keV}> \epsilon_{\rm F}$) are met. Lifting the degeneracy leads to "deflagration" rather than the runaway burning instability (the ``detonation" that plagues standard explanations of such exotic WD is effectively shut-off).  
  For the stable carbon burning case, the WD mass is below the minimum mass ($\sim 0.7 M_{\odot}$) required to sustain carbon burning in chemically homogenous main-sequence carbon stars \citep{deinzer65}.  Instead, in our model we expect the WD to puff up quickly before it cools and shrinks back to a degenerate configuration forming the iron-rich WD. During burning, a 0.575 $M_{\odot}$ WD  for example will puff up to a radius  of $\sim 3.5\times 10^9\ {\rm cm}$ from an initial radius of  $\sim 1.1\times 10^9\ {\rm cm}$ thus more than tripling its size. The WD final radius is slightly below the maximum separation that allows for carbon burning (Eq. \ref{amax}), and therefore the QS may or may not be engulfed by the puffed up WD. If the QS ends up engulfed in the puffed up star, this may lead to a phenomena ``a la Thorne-$\dot{Z}$ytkow" (\cite{TZ77}).  We expect two plausible outcomes of the engulfed QS once it starts sinking into the depth of the puffed up WD: it will either become a black hole or, form a single star by coalescing with the core of the WD. This aspect of our model will be explored in detail elsewhere.  Let us simply add that  for the heaviest WD ($M_{\rm WD} > 0.72M_{\odot}$), degeneracy will not be lifted when Carbon burning begins, and a sub-Chandrasekhar Type Ia Supernovae may result.

From Eq.~\ref{eq:Textend} and Eq.~\ref{eq:rochelobe}, it is clear that the
smaller the WD separation at the time of the QN, the larger the temperature to
which it is heated. One plausible maximum temperature the WD can be heated to is $\sim$ 134 keV, when the separation equals $a_{\rm RL}$ given in Eq.~\ref{eq:rochelobe}. This is for a CO WD with mass of about $0.6M_\odot$. 
Although not hot enough for Oxygen burning, Neon burning can occur, therefore
after burning ceases the WD will be composed of a mix of Oxygen, Neon and Magnesium. Another possibility arises if the WD is overflowing its Roche lobe, and the added mass to the NS triggers the QN. Due to the small separation, even higher WD temperatures may be reached, enabling Oxygen burning and subsequently even nuclear statistical burning and the formation of Iron group elements. The caveat to this is that if the mass ratio $M_{\rm WD}/M_{\rm NS}<2/3$, little mass is transferred - thus, the NS is less likely to undergo a QN during the mass transfer process since it may be too light to begin with. 

In cases where the QN explosion does occur, it may be asymmetric, leading to a disruption of the binary and hence allowing the heavy element WD to survive as an isolated object. In summary, we predict that:

 \begin{itemize}
  
   \item The rate of formation of these exotic WDs should be related
   to LMXBs formation rate and should be more common in LMXB formation sites.

  \item If  the parent binary survived disruption (see Table 1), \wdsrc~  should be in orbit around a QS.
   The QS X-ray luminosity is  $L_{\rm QS, X}\sim 2\times 10^{34}\ {\rm erg\ s}^{-1}\times \dot{P}_{-11}^2$ where the QS period derivative $\dot{P}$ is given in units of $10^{-11}$ s s$^{-1}$ (\cite{ouyed07}).
   
   \item  For the heaviest WD ($M_{\rm WD} > 0.72M_{\odot}$) experiencing the QN shock, degeneracy will not be lifted when Carbon burning begins, and a sub-Chandrasekhar Type Ia Supernovae may result in our model.

 \end{itemize}

\begin{table*}
\begin{center}
{\small
\begin{tabular}{|c|c|c|c|c|c|c|}\hline
$M_{\rm WD}$  & $a_{10}$ & $T_{\rm WD} $ & Nuclear burning$^{*}$ & Products &  Ablation & Binary Disruption \\\hline
  $< 0.43$ & $ < a_{10}^{\rm max} $ & $   >  8.6\  {\rm keV}$ & partial/complete $\alpha$-burning  & $\alpha$-elements & Yes & Yes (isolated QS)\\\hline
  0.43$-$0.72 & $< a_{10}^{\rm max}$ &  $   >  70.0\  {\rm keV}$ & {C\&Ne burning} & {O,Ne,Mg} & No & No/Yes$^{\dagger}$ \\\hline
  0.43$-$0.72 & $> a_{10}^{\rm max}$ & $   <   70.0\ {\rm keV} $ &Not possible&  None & No & No (X-ray hot WD) \\\hline
  $>0.72$ & $< a_{10}^{\rm max}$ &  $   >  70.0\  {\rm keV}$ &
  {Degenerate C\&Ne burning} & {O,Ne,Mg} & Yes & Yes (Isolated QS) \\\hline
\end{tabular}\\
 $^{*}$ For a CO WD, in cases when $a<a_{\rm RL}$ (see Eq.\ref{eq:rochelobe}), Oxygen burning and possible NSE can lead to nuclear burning all the way to Fe-group elements.\\
$^{\dagger}$ The velocity impacted by the QN ejecta to the WD is negligible.    
    However kicks from an asymmetric QN explosion might enhance or suppress the disruption depending on geometry. Thus, isolated $\alpha$-elements-rich WDs are not entirely ruled out in our model.\\   
    }
     \end{center}
\end{table*}

\section{A case for \grbsrc}
 \label{sec:grb11}

GRB 110328A/Swift J164449.3$+$57345 (\cite{cummings11})  has been localized  to the core of a small galaxy in the constellation Draco at a redshift of z=0.35 (\cite{levan11}) and it has remained a strong, flaring X-ray source many days after the trigger, unlike known GRBs which do not exceed a few hours of activity. The energetics of the continuing event amount to an average luminosity in the X-rays of
 $L_{\rm X} \sim 2.5 \times 10^{47}$ erg s$^{-1}$ continuing for $\sim 10^5$ s implying a total energy output of $E_{\rm X,tot} \sim 2.5\times 10^{52}$ erg (\cite{bloom11a}).   
 
 Assuming this energy is liberated in an accretion process at 10\% efficiency, the total mass involved in accretion over the first day is  $\sim 0.1 M_{\odot}$.  This led to the suggestion that the event is related to tidal disruption of a $\sim 0.5M_{\odot}$ main sequence star by a massive black hole (a few $\times  10^6 M_{\odot}$) residing in the core of the galaxy  (e.g. \cite{bloom11a}, \cite{burrows11}).  
  Here we offer an alternate explanation and argue that this event might be related
  to ablated WD material falling onto the QS. As shown next, our model comes
  with specific predictions that can be tested against observations.

\subsection{\grbsrc ~ in our model}

     In our model, LMXBs with $M_{\rm WD} < 0.43M_{\odot}$ (i.e. a He WD) experiencing the
     QN when $a_{10} < a_{10}^{\rm max}$ lead to ablation. The fate of the
     ablated material depends on many factors but here we consider a
     simplified picture. The velocity of the ablated WD material, which
     acquires momentum from the impacting QN ejecta will consist of a
     combination of:
     (i) orbital speed perpendicular to the line linking the explosion center to the WD; 
     (ii) impact velocity which is parallel to it along the radial direction;   
     (iii) ejection velocity from ablation, $v_{\rm nuc., ejec.}$, which is radial in the frame moving with the WD.

   Momentum conservation $\beta_{\rm QN}\Gamma_{\rm QN} M_{\rm QN}c 
\times (R_{\rm WD}^2/4 a^2)= M_{\rm WD} v_{\rm impact}$ leads to
 $v_{\rm impact}\simeq 14.3\ {\rm km~s}^{-1}\times E_{\rm QN, 52}^{\rm K}/ 
(a_{10}^{2} M_{\rm WD, 0.43}^{5/3})$; (here $\beta_{\rm QN}=1$ is the 
usual relativistic parameter). The WD orbital speed 
is $\sim 2100 \ {\rm km ~s}^{-1}\times M_{\rm QS,1.8}^{1/3}/a_{10}^{1/2}$ 
so for the impact velocity to exceed the orbital speed the condition is
  \begin{equation}
  M_{\rm WD}  <  0.02 \ M_{\odot} \ \frac{(E_{\rm QN, 52}^{\rm K})^{3/5}}{a_{\rm 10}^{9/10}M_{\rm QS, 1.8}^{1/5} }\ ,
  \end{equation}
  where $M_{\rm QS}$ is the mass of the QS (hardly different than its progenitor heavy NS)  in units of $1.8M_{\odot}$. While there remains the possibility that the ejected/ablated material could be concentrated in a cone away from the explosion point\footnote{Animation online at http://quarknova.ucalgary.ca/media/},  the above condition (note that $a_{10} < a_{10}^{\rm max}$) suggests that in most cases the impact velocity can be neglected and only orbital and ejection/ablation speed play a dominant role. Thus, the ablated material will most likely be confined to  a  ``fan"  of spread-out (and  expanding) WD ashes orbiting the QS.
  
  The amount of material that will be trapped by the Bondi radius (material 
heading towards the QS) is $M_{\rm trap}$ = $\Omega_{\rm trap} M_{\rm WD}$ 
where the solid angle $\Omega_{\rm trap}$ = $R_{\rm B}^2 / 4 a^2$ with 
$R_{\rm B}$ = $2GM_{\rm QS}/v_{\rm nuc.,ejec.}^2$ = $9.9\times 10^9 {\rm cm} 
\times M_{\rm QS,1.8}/\eta_{0.1}\Delta_{0.5}$ being the Bondi radius  (\cite{bondi44}). 
So the amount of material trapped by the QS is
\begin{equation}
M_{\rm trap} = Ê0.25 M_{\rm WD}\times M_{\rm
QS,1.8}^2/(\eta_{0.1}^2\Delta_{0.5}^2 a_{10}^2) \ .
\end{equation}
I.e. at least 1/4th of the WD mass will fall onto the QS.

Bondi accretion rate is  
$\dot{m}_{\rm B}= 4\pi \rho_{\rm abl}G^2 M_{\rm QS}^2/v_{\rm nuc.,ejec.}^3
\simeq  10^{-3} M_{\odot}$ s$^{-1}\times M_{\rm WD, 0.25}M_{\rm QS, 1.8}^2/
(a_{10}^3 \eta_{0.1}^{3/2}\Delta_{0.5}^{3/2})$
where  $\rho_{\rm abl}= M_{\rm WD}/(4\pi/3\times a^3)$ is the density of the falling ashes of the ablated WD. 
As shown in  \cite{romanova03},  only a fraction 
of the Bondi flux will accrete onto a magnetized, rotating compact star.
 In the non-rotating case for example (\cite{toropina03}), 
 the accretion rate is  $\dot{m}_{\rm acc.}/\dot{m}_{\rm B}\sim 6.5\times 10^{-2} \times (10 R_{\rm QS}/R_{\rm A})^5$ where $R_{\rm A}$ is the QS's Alfv\'en radius.  This gives a fiducial accretion rate of $\sim   5\times 10^{-5} M_{\odot}\ {\rm s}^{-1}\ \times (10 R_{\rm QS}/R_{\rm A})^5$.
The observed accretion X-ray luminosity of $\sim 10^{47}$ erg s$^{-1}$ suggests
$\dot{m}_{\rm acc.}\sim 10^{-6}M_{\odot}$ s$^{-1}$ which implies that $R_{\rm A} >  10 R_{\rm QS}$
 or that QS rotation is important.   We note that the corresponding Super Eddington
 accretion luminosity is justified in our model since accretion onto a QS  allows for Super Eddington
  rates\footnote{A bare QS has strong electric fields at the surface, which provides additional repulsion to any positively charged species (\cite{page02}; \cite{chengharko03}).  Magnetic field on the surface of a QS do not  change the scenario  (\cite{jaikumar06}).}. 
 
In summary, the minimum accretion energy release is  
 $\sim 0.1 M_{\rm trap}c^2\sim 2.3 \times 10^{52}$ for $M_{\rm WD}=0.25
M_\odot$.
The early, most active (i.e. spikiest), accretion phase will last for about $M_{\rm trap}/\dot{m}_{\rm acc.}\sim 2.9$ days.
 A portion of the trapped WD ashes will be expelled radially in the equatorial
 plane (\cite{toropina03}) to be accreted later.  Later accretion of non-trapped WD ashes cannot
 be ruled out either increasing the accretion activity much beyond the 3 day peak
 period. Early or late accretion of the WD ashes should 
occur along two columns  aligned with the magnetic axis of the QS (\cite{toropina03,romanova03}).  
Numerical simulations are required to better quantify the numbers and accretion
 geometry in our model. Some predictions are:

 \begin{itemize}
 
  \item   \grbsrc~ might  be a short hard GRB as described in our model (see OSJ).
  This could be tested by signatures of an EE in the 
   earliest observations  of this event before the accretion phase.
   
   \item Type I SN-like
   features from the non-trapped (ablated) WD ashes should also be distinguishable (see \S 5 in OSJ). 
 
 \item During the accretion phase, we expect signatures of $\alpha$-elements in this
 system since the infalling material is made of  ashes from the burnt and ablated WD.
 
 \item  If the QS does not turn into a BH following the accretion phase, an X-ray
 plateau is expected from the QS spin-down luminosity.
 
 \item   If  the QS turns into a black hole (if it exceeds its mass limit), 
  additional flaring activity during or towards the end of the accretion phase should
   be evident. This could be seen as a jump (decrease or increase) in the X-ray activity caused by the change in accretion efficiency as the QS turns into a BH.

\end{itemize}

\section{Discussion and Conclusion}

In this paper, we have relied on order of magnitude estimates to present a plausible model for the flaring in GRB 110328 A and
the occurrence of low-mass Iron-rich WDs. To compare to detailed observations, it is necessary to perform detailed
multi-dimensional, hydrodynamical simulations of the explosion of a QN in a LMXB, and couple
the results with a nuclear network code to properly capture the relevant nucleosynthesis
 during WD burning.  Both these areas are works in progress (Niebergal et al. (2010) and \cite{camille11} respectively). 
 In the meantime, we have presented some broad testable predictions of our model that can be checked by continued observations.

As already argued in OSJ, our model suggest that the GRB engine for some short GRBs resides in globular clusters (GCs) where many LMXBs are seemingly found  \citep{bogdabov06,camilo05}.  Thus,  
 events such as GRB 110328 A  and $\alpha$-elements WDs
in our models will most likely reside in GCs and as such we expect them to be astrometrically coincident with halos and nuclei of  galaxies.   

We have discussed how a QN may be able to restart fusion processes in a
WD that happens to be in a close binary with the exploding NS, which could
explain the occurrence of heavy elements in WDs. We found that WD more
massive than $0.43 M_\odot$ can burn carbon without being ablated. One may wonder if a QN is really necessary, or if a type 1a supernova
occuring with a WD orbiting it could produce similar events? 
However, the kinetic energy of the SN ejecta in a type
1a SN is thought to be about $10^{51}$ erg (rather than $10^{52}$ erg as in
the case of a QN), making it unlikely that the temperature will be high
enough to start nuclear processes. Finally, we note that in the cases when the WD is not heated to temperatures that allow for nuclear burning, it may still get very hot. This therefore opens up the possibility of a very old and hot WD.

To summarize: If a QN occurs in a close binary with a WD, then the QN may
trigger nuclear burning in the WD. If the WD is an He WD or a very low mass CO
WD ($M<0.43 M_\odot$), then the WD will get completely ablated and a GRB 110328A phenomena may result. If the CO WD has a higher mass it may survive the QN,
and the nuclear processes will lead to the formation of a heavy element WD.
Detailed hydrodynamical and nucleosynthesis calculations
are ultimately required to test our assumption that the WD can, under certain conditions, avoid the thermonuclear runaway associated with type Ia SN explosions.


\begin{acknowledgements}
The research of R. O. is supported by an operating grant from the
National Science and Engineering Research Council of Canada (NSERC). P. J.
acknowledges start-up funds from California State University Long Beach. 
This work has been supported, in part, by grant AST-0708551
from the U.S. National Science Foundation and, in part, by
grant NNX10AC72G from NASA's ATP program.
\end{acknowledgements}

\end{document}